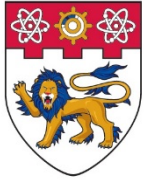

# Sustainable Development Goals (SDGs): New Zealand's Outlook with Central Bank Digital Currency and SDG 8 Realization on the Horizon

Qionghua Chu
Nanyang Business School, Nanyang Technological University

# Sustainable Development Goals (SDGs): New Zealand's Outlook with Central Bank Digital Currency and SDG 8 Realization on the Horizon

By Qionghua Chu (Ruihua, Katherine), CAIA, FRM

**Abstract**

Central Bank Digital Currency (CBDC) may assist New Zealand accomplish SDG 8. I aim to evaluate if SDGs could be achieved together because of mutual interactions between SDG 8 and other SDGs. The SDGs are categorized by their shared qualities to affect and effect SDG 8. Also, additional SDGs may help each other achieve. Considering the CBDC as a fundamental stimulus to achieving decent work and economic growth, detailed study and analysis of mutual interactions suggests that SDG 8 and other SDGs can be achieved.

## 1. Introduction

SDG 8 – decent work and economic growth – and sixteen others are scheduled for 2030 by the UN. Since goals interact and the Reserve Bank of New Zealand (RBNZ)'s dual objectives of long-run average inflation target midpoint of 2% and maximum sustainable employment align with SDG 8, realizing the goals through CBDC issuance (RBNZ, 2021), is worth considering.

## 2. Impacts of SDG 8 Realization and CBDC Issuance on Other SDGs

Despite the inverted yield curve, the CBDC's potential benefits to SDG 8 may help achieve other targets. As I demonstrate in Figure 3, SDG 8 could help fulfill other goals because the economy may provide funding and resources, and while other goals progress, they create a healthy ecosystem to help SDG 8 be realized more efficiently and effectively in NZ. The mutual interactions emphasize the importance of achieving SDG 8 with the CBDC in mind while also achieving other objectives to mutually support each other.

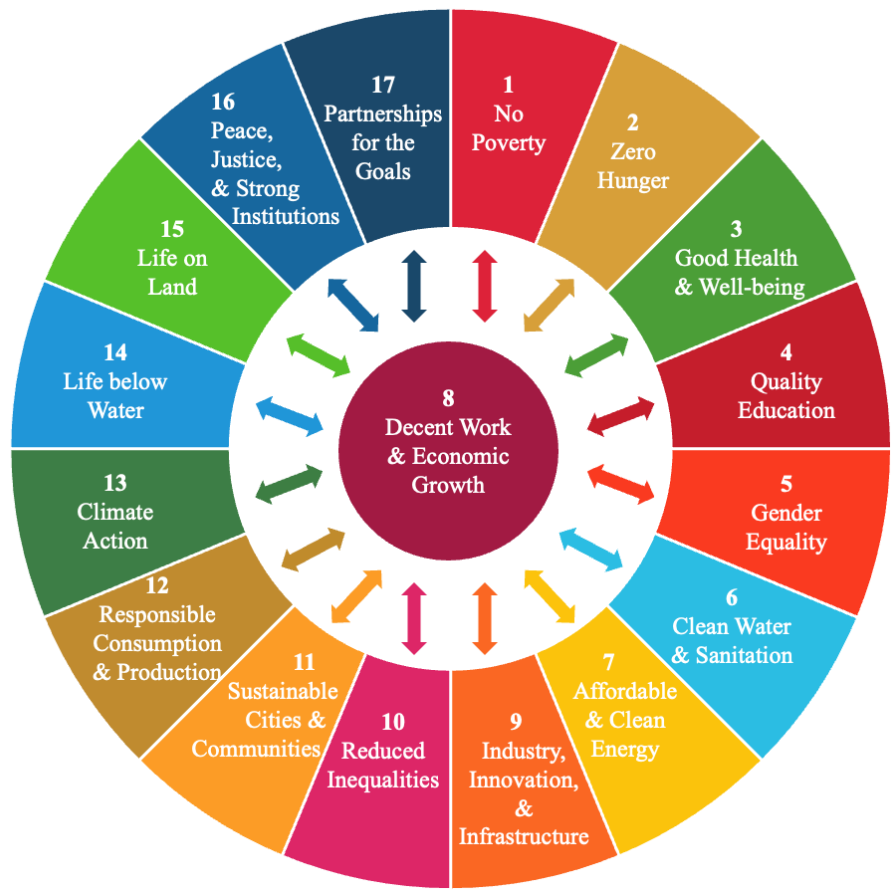

**Figure 3**. *Interactions of SDG 8 with Others*

On SDG 1, no poverty, 2, zero hunger, 3, good health and well-being, and 4 quality education, in the year ended June 2021, 13.6% of children (156,700) lived in households with less than 50% median equivalized disposable income before taking away housing costs, a 17.6% decrease from 16.5% (183,400) in June 2018 (SNZ, 2022). The CBDC and SDG 8 may increase disposable income for those households by providing more education, training, and employment in high-value-added and less labor-intensive sectors related to hardware and software infrastructures of the CBDC and its associated industries. More households out of the less than 50% median equivalized disposable income before housing costs may create a virtuous cycle that creates more decent jobs. More consumption from those households may boost corporate investments, creating more quality jobs and economic growth. With less poverty, SDG 2 of zero hunger may be achieved. 14.9% of 4- to 5-year-olds in NZ are obese, with 2.9% severe (Butler et al., 2021). This is malnutrition. Instead of junk food, children might eat fresh fruits and veggies with more respectable work, economic prosperity, and reduced poverty. With more children needing nutritious food, business investments in and demand for workers in food and beverages, farming, rural infrastructures, food manufacturing, transportation, and other related industries may also rise, leading to more decent work and economic growth to meet SDG 8. While SDG 2 is met, SDG 3 of good health and well-being may be too. Improved nutrition and healthcare have reduced NZ's under-five death rate from 6.3% in 2009 to 4.7% in 2021 (UNICEF, n.d.). As good jobs and economic growth increase, government money can be efficiently directed to healthcare and other areas like transportation, reducing mortality rates from diverse causes including road traffic accidents. Workers may be more fit to work when general health and well-being improve, increasing disposable household income. More individuals are eager and able to consume, which boosts business, investments, and job development. This will support SDG 8 of decent jobs and economic growth. SDG 3 also ensures the young are healthy enough for SDG 4 – decent education. Healthier children can focus on early childhood education and meet minimum reading and math standards. As they grow up, a better foundation and quality education could qualify them for decent work and boost economic growth. SDG 8 could help achieve SDGs 1, 2, 3, and 4, and the latter could aid the former.

As for equalities, SDG 5, gender equality, and SDG 10, reduced inequalities, as SDG 8 aims to create decent work and economic growth, and the high-valued added and less labor-intensive jobs might be created with CBDC issuance on its own and in related industries. This could potentially enable gender equality and reduce inequalities. While NZ has enacted the Equal Pay Amendment

Act 2020 and the national pay gap has reduced from 16.3% to 9.2% for 25 years until 2022, the worrisome slowing progress from 2017 to 2022 (NZ Parliament, 2022) suggests more efforts might be needed. On SDG 5, as women and girls receive more education and training to prepare for and embrace decent work, their increasing ability to become financially independent could potentially help them to reduce different forms of discrimination due to their gender. Besides, as they have more equal status, they might be able to contribute to economic growth by increasing disposable household income, consumption, and business investments, and creating more decent work opportunities. Additionally, on SDG 10, with more decent work and economic growth attributed to the potential issuance of the CBDC, the inequalities faced by different genders, age groups, or people with disabilities might be reduced. For example, the average household equivalized disposable income for disabled people of NZ$42,239 is 18.27% lower than that of the non-disabled of NZ$51,683 for the year ended June 2021 (SNZ, 2022). With high value-added and less labor-intensive jobs created along the process of fulfilling SDG 8 via the CBDC potential issuance, the disabled people could receive more education and training to qualify for more decent work, increase disposable household income, and propel economic growth. Therefore, the process of realizing SDG 8 with the CBDC potential issuance could help to achieve SDGs 5 and 10, and in turn, help to fulfill SDG 8.

For SDGs 6, 13, 14, and 15 of clean water and sanitation, climate action, life below water, and life on land, in fulfilling SDG 8, as the economy improves with decent work and economic growth, more resources may be available to set up and improve infrastructures for SDG 6 – clean water and sanitation, climate action, life below water, and life on land. While the number of people using securely managed drinking water services has improved from 82.0% in 2000 to 100.0% in 2020, the reduction in safely treated household wastewater flow from 85.14% in 2020 to 84.48% in 2022 (UN, n.d.) is concerning. With more good work and economic growth, and maybe the CBDC, more government budgets may be available for wastewater treatment to meet clean water and sanitation needs. Better living conditions will allow more people to focus on training and education and be better prepared to take decent jobs and boost economic growth as water-use efficiency and integrated water resources management for safely managed sanitation services improve. On SDG 13 of climate action, NZ has 100% local governments that have adopted and implemented disaster risk reduction strategies to align with national strategies, but 2,053 people were affected by a disaster in 2021, up 175.57% from 745 in 2020 (UN, n.d.). With the development of SDG 8 and better economic conditions, more resources could be available to build stronger disaster-prevention, detection, and mitigation infrastructure, accomplishing SDG 13 of combating climate change. With better SDG 13, the average Marine Key Biodiversity Areas (KBAs) covered by protected areas may increase from 42.62% in 2000 to 46.54% in 2022 (UN, n.d.) to better protect, conserve, and sustainably use seas, oceans, and marine resources on SDG 14 of life below water. Better environmental conditions could aid SDG 8 as maritime resources are preserved. Since offshore aquaculture, minerals, and fishers fell 14.44% from 2020 to 2021, the marine economy's GDP contribution dropped to NZ$3.91 billion (SNZ, 2023), fulfilling SDG 8 could boost economic growth, government revenue, and marine ecosystem sustainability. A stronger marine ecosystem could boost economic growth and decent work. On SDG 15, Mountain KBAs covered by protected areas increased from 27.84% in 2000 to 34.02% in 2022 (UN, n.d.), but more decent work and economic growth could allow the government to allocate more resources to protect and conserve terrestrial ecosystems and sustainably develop land resources. Conversely, reversed land degradation, combatted desertification, and better forest management could lead to more decent terrestrial ecosystem conservation and sustainable economic growth. Thus, while SDG 8 may aid SDGs 6, 13, 14, and 15, the latter may also aid the former.

With regards to SDGs 7 of affordable and clean energy, 9 of industry, innovation, and infrastructures, 11 of sustainable cities and communities, and 12 of responsible consumption and production, SDG 8 is quintessential. NZ has already performed well for SDG 7, with reliance on clean fuels and technology at 95.00% in 2021 (UN, n.d.), electricity provision from renewable energy increasing from 81.1% in 2020 to 82.1% in 2021, and the share of renewable energy supply to total reaching its highest level of 40.8% (Ministry of Business, Innovation and Employment, 2022). With more decent work and economic growth, more resources could be allocated to infrastructure. Additionally, more efficient, clean, and economical energy generation could enable decent work on hardware infrastructures and software monitoring to support economic growth and decent work. On SDG 9, the density of full-time researchers has increased by 121.49% from 2,643 per million people in 2000 to 5,854 per million in 2019 (UN, n.d.). With the CBDC to boost decent work and economic growth, the rise might be greater. In addition, the CBDC may demand additional R&D in its own and associated businesses to construct hardware and software infrastructures, raising R&D expenditure as a proportion of GDP from 1.40% in 2019 (UN, n.d.).

More full-time researchers and R&D could lead to breakthroughs in high-value, low-labor industries, improving jobs and economic growth. Thus, SDGs 7, 8, and 9 could support one other.

SDG 8 could help accomplish SDGs 16 (peace, justice, and strong institutions) and 17 (partnerships for the goals). NZ did well on SDG 16, with the 12-month prevalence of physical violence falling from 5.7% in 2014 to 2.5% in 2021 (UN, n.d.). However, with SDG 8 and more decent jobs and economic growth, disadvantaged people may be able to get skills and education to become financially independent and protect themselves from physical assault. Economic growth may also give peace and justice organizations additional financing and resources for sustainable development. As society grows more peaceful and inclusive, it may be easier for people to find good jobs and boost economic growth. On SDG 17, NZ has strengthened both domestic and international efforts to revitalize global partnerships for sustainable development, as shown by the increase in total government revenue as a percentage of GDP from 28.90% in 2000 to 36.98% in 2021 and the rise of 137.44% in net Official Development Assistance (ODA) from US$253.89 million to US$602.83 million (UN, n.d.). However, more could be done to promote decent work and economic growth. With additional cash, the NZ government can apply sustainable development concepts locally and through net ODA to support poor nations and LDCs. As local and foreign economies grow sustainably, their environments become more conducive to decent employment and economic progress. Thus, SDG 8 can aid SDGs 16 and 17, and the latter can aid the former.

# 3. Conclusion and Policy Considerations

In a nutshell, notwithstanding the inverted yield curve, I firmly believe that CBDC issuance in NZ could enhance economic development and provide decent jobs to fulfill SDG 8. Moreover, although reaching all 17 SDGs by 2030 may seem difficult, focusing on SDG 8 via CBDC issuance may make the other 16 SDGs easier to achieve.

Since the twelve goals are interrelated, policymakers may issue CBDC to achieve some or all. When the yield curve inverts, Targets 8.1, 8.2, 8.3, 8.4, 8.5, and 8.6 can boost discretionary household income and job prospects to sustain consumption, business investment, and hence, economic growth. In the medium to long run, yield curve normalization will boost CBDC issuance. With CBDC issuance, increased government spending on law enforcement and rights education could accomplish Targets 8.7 and 8.8, while sustainable tourism infrastructure, easier financial services, more AFT resources, and better youth employment outlook could advance Targets 8.9, 8.10, 8.a, and 8.b. Lastly, authorities can collaborate with experienced global partners to construct CBDC and achieve SDG 8.

Moreover, as SDG 8 and other goals are interrelated, policymakers may consider adopting the CBDC with decent work, economic growth, and other factors in mind to meet SDGs holistically. Policymakers in various institutions can work on how to implement the CBDC in the financial and related sectors to set up hardware and software infrastructure for sustainable decent work, economic growth, and related goals. Another option is for policymakers to create a roadmap of goals to achieve first with the CBDC in mind to achieve the SDGs efficiently and effectively. Economic growth may directly affect SDGs 1, 2, 3, and 4, therefore it can be achieved first to lay the groundwork for subsequent objectives. Finally, NZ policymakers may collaborate with international partners to study the CBDC's pros and downsides. Knowing how to modify the digital NZD to match unique home contexts and global trends may be an essential policy consideration to efficiently fulfill SDGs.

Despite CBDC issuance risks, I envision bright prospects for SDG 8 to be realized in NZ.

# References

Butler, É.M., Pillai, A., Morton, S.M.B. et al. A prediction model for childhood obesity in New Zealand. *Scientific Reports*, *11*(1), 6380.

New Zealand Parliament. (2022, October 7). Fifty years of the Equal Pay Act 1972. https://www.parliament.nz/en/pb/library-research-papers/research-papers/fifty-years-of-the-equal-pay-act-1972/#:~:text=The%20Equal%20Pay%20Act%201972%20grants%20equal%20pay%20for%20equal,Service%20Equal%20Pay%20Act%201960

Reserve Bank of New Zealand. (2021). *The future of money – Central Bank Digital Currency*. Reserve Bank of New Zealand. https://www.rbnz.govt.nz/-/media/project/sites/rbnz/files/consultations/banks/future-of-money/cbdc-issues-paper.pdf

Reserve Bank of New Zealand. (2023). Monetary policy. https://www.rbnz.govt.nz/monetary-policy

Statistics New Zealand. (2022, February 24). Child poverty statistics show all measures trending downwards over the last three years. https://www.stats.govt.nz/news/child-poverty-statistics-show-all-measures-trending-downwards-over-the-last-three-years

Statistics New Zealand. (2022, February 24). Household income and housing-cost statistics: Year ended June 2021. https://www.stats.govt.nz/information-releases/household-income-and-housing-cost-statistics-year-ended-june-2021/

Statistics New Zealand. (2023, March 7). Environmental-economic accounts: Data to 2021. https://www.stats.govt.nz/information-releases/environmental-economic-accounts-data-to-2021/

United Nations. (n.d.). SDG country profile: New Zealand. Retrieved May 12, 2023, from https://unstats.un.org/sdgs/dataportal/countryprofiles/nzl#goal-8

United Nations International Children's Emergency Fund. (n.d.). UNICEF data warehouse: Indicators percentage of children (aged 5-17 years) engaged in child labor (economic activities) and percentage of children (aged 5-17 years) engaged in child labor (economic activities and household chores). Retrieved May 12, 2023, from https://data.unicef.org/resources/data_explorer/unicef_f/?ag=UNICEF&df=GLOBAL_DATAFLOW&ver=1.0&dq=.PT_CHLD_5-17_LBR_ECON+PT_CHLD_5-17_LBR_ECON-HC..&startPeriod=2016&endPeriod=2022